\documentclass[twocolumn,showpacs,preprintnumbers,amsmath,amssymb]{revtex4}
\preprint{}
\usepackage{graphicx}
\usepackage{dcolumn}
\usepackage{bm}
\usepackage{mathrsfs}
\begin{document}
\title{ Degree of Fuzziness in Coarsened Measurement References }
\author{Dong  Xie}
\email{xiedong@mail.ustc.edu.cn}
\author{An Min Wang}
 \email{anmwang@ustc.edu.cn}
  \affiliation{Department of Modern Physics , University of Science and Technology of China, Hefei, Anhui, China.}
\begin{abstract}
It has been found that the quantum-to-classical transition can be observed independent of macroscopicity of the quantum state for a fixed degree of fuzziness in the coarsened references of measurements. Here, a general situation, that is the degree of fuzziness can change with the rotation angle between two states (different rotation angles represent different references), is researched based on the reason that the fuzziness of reference can come from two kinds: the Hamiltonian (rotation frequency) and the timing (rotation time). Our results show that, for the fuzziness of Hamiltonian alone, the degree of fuzziness for reference will change with the rotation angle between two states and the quantum effects can still be observed no matter how much degree of fuzziness of Hamiltonian; for the fuzziness of timing, the degree of coarsening reference is unchanged with the rotation angle. Moreover, during the rotation of the measurement axis, the decoherence environment can also help the classical-to-quantum transition due to changing the direction of measurement axis.

\end{abstract}

\maketitle

\section{Introduction}
Since the quantum phenomena were observed on a microscopic scale, the
quantum-to-classical transition has attracted a lot of attentions. As we all
known, two crucial elements in the framework of quantum mechanics: one is
the state of a physical system represented by a wave function, and another
is measurement represented by non-negative operators. Moreover, the
decoherence of a system from quantum state to the classical one is the main
reason that the macroscopic world is classical \cite{lab1, lab2}, not quantum, and the
decohenrence happens due to its unavoidable interactions with environments.

On the other hand, an explanation of the quantum-to-classical transition comes from the measurement.  J. Kofler and $\check{C}$. Brukner \cite{lab3} firstly attributed coarsening of measurements to the cause. Then, a lot of works about the influence of coarsening of measurements on the quantum-to-classical transition \cite{lab4,lab5,lab6,lab7,lab8,lab9,lab10,lab11,lab12,lab13} appeared. They mainly focused on coarsening the measuring resolution in the final detection.
There are two steps for a complete measurement process: the first step is to set a measurement reference and control it, and the second step is the final detection with the corresponding projection operator. Recently, Hyunseok Jeong, et.al \cite{lab14} shed light upon the appearance of a classical world from another angle: coarsening the measurement reference. In their scheme, the control of
the measurement reference is described by an appropriate
unitary operator with a reference variable applied to the projection operator, and they considered the fixed degree of fuzziness in the coarsened references of measurements, which didn't change with the rotation angle of the measurement axis corresponding to the different unitary operations.

In this article, we consider the general situation: the degree of fuzziness can change in different coarsened reference based on the fact that the fuzziness of coarsened reference comes from the fuzziness of Hamiltonian ( rotation frequencies of the reference) and timing (when to do the final detection). It is found that for only the fuzziness of Hamiltonian, the degree of fuzziness in the coarsening reference measurement increases with the rotation angle, while for only fuzziness of timing, the degree of fuzziness is unchanged with the rotation angle. Further, we show that the Bell function $|B|$ decreases with the fuzziness, when $|B|\geq2$; when $|B|<2$, it can increase with the fuzziness. And we study other ways to detect the quantum effect ($B>2$), such as changing the angle, frequency, and steps of rotation. Moreover, we obtain the conclusion that the decoherence environments can help to detect the quantum effect in the coarsening reference during the rotation of reference. Our results will further shed light on the fuzziness in coarsening reference measurements.

The rest of this article is arranged as follows. In the Sec.II, we classify the fuzziness of coarsening reference as three kinds of fuzziness: only Hamiltonian (section A), only timing (section B), and both of Hamiltonian and timing (section C). In the Sec.III, the decoherence environment is studied in the coarsening reference measurements. Finally, we make a conclusion in the sec.III.

\section{ three kinds of fuzziness}
Consider the generic example: an infinite dimensional system with the orthonormal basis basis set ${|o_n\rangle}_{n=-\infty}^{n=\infty}$. The dichotomic measurement $O^k$ with eigenvalues $\pm1$ can be given by
\begin{equation}
O=O_+-O_-,
\end{equation}
in which,
\begin{equation}
O_+=\sum_{n=1}^{\infty}|o_n\rangle\langle o_n|, O_-=\sum_{n=-\infty}^{0}|o_n\rangle\langle o_n|.
\end{equation}

Let us consider a type of entanglement as follows:
\begin{equation}
|E_n\rangle=\frac{1}{\sqrt{2}}(|o_n\rangle|o_{-n}\rangle+|o_{-n}\rangle|o_n\rangle).
\end{equation}
A unitary transform $U(\theta)$ is described by
\begin{equation}
\begin{split}
U(\theta)|o_n\rangle=\textmd{cos}\theta|o_n\rangle+\textmd{sin}\theta|o_{-n}\rangle,\\
U(\theta)|o_n\rangle=\textmd{sin}\theta|o_n\rangle-\textmd{cos}\theta|o_{-n}\rangle,
\end{split}
\end{equation}
where $\theta=wt$ represents the rotation angle, $w$ denotes the rotation frequency, and $t$ is the corresponding time of rotation. The coarsened version of the unitary operation applied to the projection operator $O$ can be given as
\begin{equation}
O_\Delta(\theta_0)=\int d\theta P_\Delta(\theta-\theta_0)[U^\dagger(\theta)O U(\theta)],
\end{equation}
where $P_\Delta(\theta-\theta_0)=\frac{1}{\sqrt{2\pi}\Delta }\exp[-\frac{(\theta-\theta_0)^2}{2\Delta ^2}]$ is the normalized Gaussian kernel with standard deviation $\Delta$, which quantifies the degree of fuzziness in the coarsened measurement reference.

The fuzziness of rotation angle $\theta_0=w_0t_0$ comes from the fuzziness of $w_0$ and $t_0$.
So it is necessary to explore the fuzziness of rotation angle from the following three aspects:

\subsection{Coarsening the rotation frequency $w_0$}
Due to the fuzziness of Hamiltonian which performs the unitary operation, the rotation angle is coarsened.
The corresponding fuzzy version of the rotation frequency $w_0$ can be written as a Gaussian distribution: $P_{\Delta_{w_0}}(w-w_0)=\frac{1}{\sqrt{2\pi}{\Delta_{w_0}} }\exp[-\frac{(w-w_0)^2}{2{\Delta ^2_{w_0}}}]$.

Then, the coarsened version of the unitary operation can be described as
\begin{equation}
\begin{split}
O_{\Delta_{\theta_0}}(w_0t)=\int dw P_{\Delta_{w_0}}(w-w_0)[U^\dagger(wt)OU(wt)]\\
=\int d\theta P_{\Delta_{\theta_0}}(\theta-\theta_0)[U^\dagger(\theta)O U(\theta)],
\end{split}
\end{equation}
where the rotation angle $\theta_0=w_0t$, and $\Delta_{\theta_0}=\Delta_{w_0}\theta_0/w_0$. It means that the degree of fuzziness in coarsening the measurement reference increases with the rotation angle $\theta_0$.

Let us consider to rotate the reference with many steps. For example, one first rotates the reference with the angle $\theta_0/2$. Following that, rotate the reference with another angle $\theta_0/2$. The reference is rotated with angle $\theta_0$ in total. But, due to the two times adjustment of unitary operation, the degree of fuzziness $\Delta_{\theta_0}|_{0\rightarrow\theta_0/2\rightarrow\theta_0}$ will become different with a single rotation $\Delta_{\theta_0}|_{0\rightarrow\theta_0}$. Applied to the projection operator $O$,
\begin{equation}
\begin{split}
O_{\Delta_{\theta_0}}(w_0t)|_{0\rightarrow\theta_0/2\rightarrow\theta_0}=
\int dw'\int dwP_{\Delta_{w_0}}(w'-w_0)\\ P_{\Delta_{w_0}}(w-w_0)[U^\dagger(\frac{w+w'}{2}t)OU(\frac{w+w'}{2}t)]\\
=\int d\theta P_{\Delta_{\theta_0}|_{0\rightarrow\theta_0/2\rightarrow\theta_0}}(\theta-\theta_0)[U^\dagger(\theta)O U(\theta)],
\end{split}
\end{equation}
where the degree of fuzziness $\Delta_{\theta_0}|_{0\rightarrow\theta_0/2\rightarrow\theta_0}=\Delta_{\theta_0}|_{0\rightarrow\theta_0}/\sqrt{2}$.
If one performs $N$ rotations, the degree of fuzziness  $\Delta_{\theta_0}|_{0\rightarrow\theta_0/N\rightarrow2\theta_0/N\rightarrow...\theta_0}=\Delta_{\theta_0}|_{0\rightarrow\theta_0}/\sqrt{N}$.
So it signifies that many steps of rotation can reduce the fuzziness of coarsening the measurement reference.

The correlation function is the expectation value of the measurement operators as
\begin{equation}
E_{\Delta_a,\Delta_b}(\theta_a,\theta_b)=\langle O_{\Delta_a}(\theta_a)\otimes O_{\Delta_b}(\theta_b)\rangle_{ab},
\end{equation}
where the average is taken over entangled state $|E_n\rangle_{ab}$ in Eq.(2) and $\Delta_a,\Delta_b$ are the corresponding standard deviations for rotation angle $\theta_a$ and $\theta_b$.
Then, the Bell function \cite{lab15,lab16} can be obtained as
\begin{equation}
\begin{split}
B=E_{\Delta_{a1},\Delta_{b1}}(\theta_a,\theta_b)+E_{\Delta'_{a2},\Delta_{b2}}(\theta'_a,\theta_b)\\
+E_{\Delta_{a3},\Delta'_{b3}}(\theta_a,\theta'_b)-E_{\Delta'_{a4},\Delta'_{b4}}(\theta'_a,\theta'_b),
\end{split}
\end{equation}
where the subscript $i=1,2,3,4$ distinguishes the degree of fuzziness $\Delta$ for the same rotation angle $\theta$  in different joint measurements. Choose the value of parameter: $\Delta_{xi}=0$ for $x=a,b$ and $i=1,2,3$ ( the corresponding rotation frequency $w_0$ close to infinity); $\Delta_{a4}=0$ or $\Delta_{a4}=0$ (corresponding frequency $w_0$ close to 0); $\theta_a=\theta_b=0$ and $\theta'_a=\theta'_b=2\pi$ .   Based on the simple calculation, we find that the maximum of Bell function $|B|\approx3>2\sqrt{2}$ for both quantum state and classical one. It means the Bell function can't show the difference between quantum and classical for changed degree of fuzziness $\Delta$. So the Bell function should be defined as
\begin{equation}
\begin{split}
B=E_{\Delta_{a},\Delta_{b}}(\theta_a,\theta_b)+E_{\Delta'_{a},\Delta_{b}}(\theta'_a,\theta_b)\\
+E_{\Delta_{a},\Delta'_{b}}(\theta_a,\theta'_b)-E_{\Delta'_{a},\Delta'_{b}}(\theta'_a,\theta'_b).
\end{split}
\end{equation}
In this definition, we obtain that as the result in the ref.\cite{lab14}, the Bell function $|B|$ decreases with the degree of fuzziness $\Delta_{w_0}$ for $|B|\geq2$. It is proved as follows:
\begin{equation}
\begin{split}
\frac{dB}{d\Delta_{w_0}}&=-2\Delta_{w_0}/w_0^2[E_{\Delta_{a},\Delta_{b}}(\theta_a,\theta_b)({\theta_a}^2+{\theta_b}^2)+\\
&E_{\Delta'_{a},\Delta_{b}}(\theta'_a,\theta_b)({\theta'_a}^2+{\theta_b}^2)+\\
&E_{\Delta_{a},\Delta'_{b}}(\theta_a,\theta'_b)({\theta_a}^2+{\theta'_b}^2)-\\
&E_{\Delta'_{a},\Delta'_{b}}(\theta'_a,\theta'_b)({\theta'_a}^2+{\theta'_b}^2)]
\end{split}
\end{equation}
To note that correlation function $|E|\leq1$ and the Bell function $|B|\geq2$, we can obtain that the differential coefficient $\frac{d|B|}{d\Delta_{w_0}}\leq0$, which signifies the Bell function $|B|$ decreases with $\Delta_{w_0}$.
\begin{figure}[h]
\includegraphics[scale=0.8]{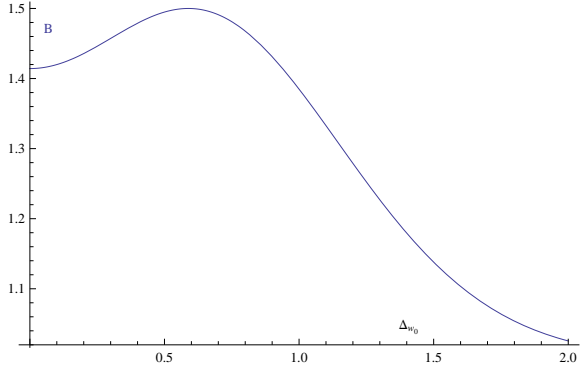}
\caption{\label{fig.2}The graph shows that the Bell function $|B|$ can increase with $\Delta_{w_0}$. Here the parameter is given as: $\theta_a=\theta_b=0$, $\theta'_a=\pi/8$, $\theta'_b=-\pi/8$, and $|w_0|=\sqrt{2}\pi/8$. }
 \end{figure}
However, when $|B|<2$, the Bell function can increase with $\Delta_{w_0}$, as shown in Fig.1. The reason is that the degree of coarsening the measurement reference changed with the rotation angle.

And we obtain that no matter how much the degree of fuzziness $\Delta_{w_0}$, the quantum effects ($|B|>2$) can be observed by reducing the rotation angle. Simply verify it: for the parameter $\theta_a=\theta_b=0$, $\theta'_a=\theta'_b=\lambda\ll1$, and $|w_0|=1$, we can derive that $B=2+4\lambda^2+\mathcal{O}(\lambda^3)>2$. Essentially, it is because that the degree of coarsening the reference increase with the rotation angle: $\Delta_{\theta_0}=\Delta_{w_0}\theta_0/w_0$.
\subsection{Coarsening the rotation time $t_0$}
When the rotation frequency $w_0$ can be controlled precisely, the fuzziness of reference can come from the coarsened timing.
Similar with the above section, the Gaussian distribution of rotation time $t$ is given by $P_{\Delta_{t_0}}(t-t_0)=\frac{1}{\sqrt{2\pi}{\Delta_{t_0}} }\exp[-\frac{(t-t_0)^2}{2{\Delta ^2_{t_0}}}]$.

Then, the coarsened version of the unitary operation can be described as
\begin{equation}
\begin{split}
O_{\Delta_{\theta_0}}(w_0t_0)=\int dt P_{\Delta_{t_0}}(t-t_0)[U^\dagger(w_0t)OU(w_0t)]\\
=\int d\theta P_{\Delta_{\theta_0}}(\theta-\theta_0)[U^\dagger(\theta)O U(\theta)],
\end{split}
\end{equation}
where the rotation angle $\theta_0=w_0t_0$, and $\Delta_{\theta_0}=\Delta_{w_0}w_0$. It means that for the fixed rotation frequency, the degree of fuzziness in coarsening the measurement reference is unchanged with the rotation angle $\theta_0$. So it is just the situation considered in the ref.\cite{lab14}. Here, the Bell function $B$ follows the definition in the Eq.(10). And it is easy to verify that the Bell function $|B|$ decreases with the degree of fuzziness $\Delta_{t_0}$.

Contrary with the situation in the above section, the many steps of rotation will increase the degree of coarsening reference: $\Delta_{\theta_0}|_{0\rightarrow\theta_0/N\rightarrow2\theta_0/N\rightarrow...\theta_0}=\sqrt{N}\Delta_{\theta_0}|_{0\rightarrow\theta_0}$, where $N$ is the steps of rotation. It is because that the many steps of rotation increase the times of timing.
\subsection{coarsening both rotation frequency and time}
The more general situation is that both the rotation  frequency and time are coarsened. Now the coarsened version of the unitary operation can be described as:
\begin{equation}
\begin{split}
&O_{\Delta_{\theta_0}}(w_0t_0)\\
&=\int dw \int dtP_{\Delta_{w_0}}(w-w_0)P_{\Delta_{t_0}}(t-t_0)[U^\dagger(wt)OU(wt)]\\
&=\int d\theta |\int dw\frac{1}{2\pi\Delta_{t_0}\Delta_{w_0}w}\exp[-\frac{(w-w_0)^2}{2\Delta^2_{w_0}}-\frac{(\theta/w-t_0)^2}{2\Delta^2_{t_0}}]|\\
&\times[U^\dagger(wt)OU(wt)]\\
&=\int d\theta P_{\Delta_{\theta_0}}(\theta-\theta_0)[U^\dagger(\theta)O U(\theta)].
\end{split}
\end{equation}
So the distribution of rotation angle is obtained:
\begin{equation}
\begin{split}
P_{\Delta_{\theta_0}}(\theta-\theta_0)&=|\int dw\frac{1}{2\pi\Delta_{t_0}\Delta_{w_0}w}\\
&\times\exp[-\frac{(w-w_0)^2}{2\Delta^2_{w_0}}-\frac{(\theta/w-t_0)^2}{2\Delta^2_{t_0}}]|.
\end{split}
\end{equation}
From Fig.2 and Fig.3, we can see that the function $P_{\Delta_{\theta_0}}(\theta-\theta_0)$ isn't a Gaussian distribution, the central value isn't at $\theta_0=w_0t_0$ as expected, and it's discontinuous at $\theta=0$. As shown in Fig.4, it is continuous only when the rotation time $t_0=0$.
\begin{figure}[h]
\includegraphics[scale=1.4]{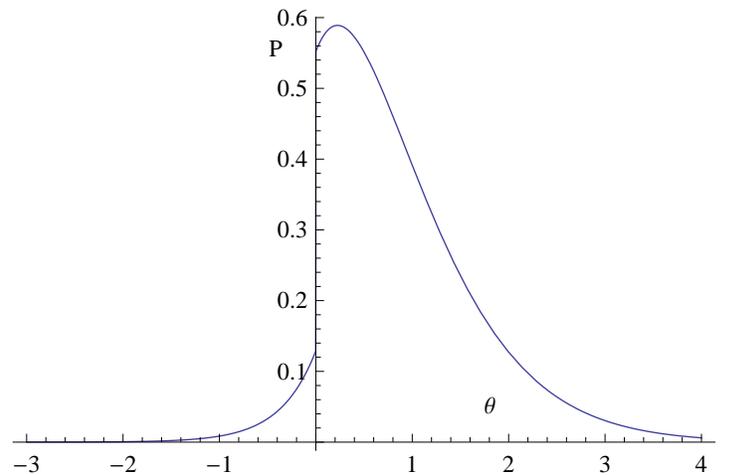}
\caption{\label{fig.2}The graph shows that the distribution $P_{\Delta_{\theta_0}}(\theta-\theta_0)$ changes with the rotation angle $\theta$, where the parameter is given by $w_0=1$, $t_0=\pi/4$, $\Delta_{w_0}=0.6$, and $\Delta_{t_0}=0.6$.}
 \end{figure}
 \begin{figure}[h]
\includegraphics[scale=1]{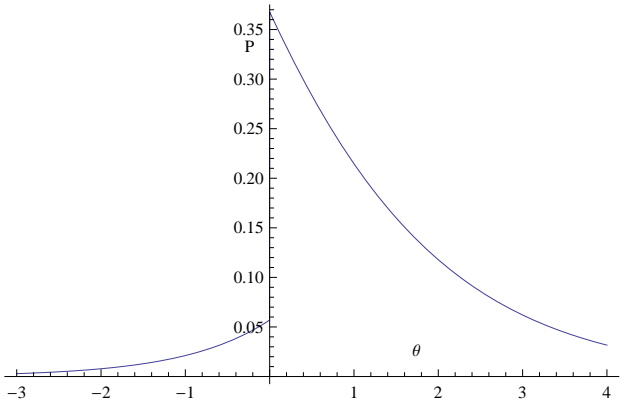}
\caption{\label{fig.2}Like the Fig.2, here we choose lager parameter values for $\Delta_{w_0}$ and $\Delta_{t_0}$: $w_0=1$, $t_0=\pi/4$, $\Delta_{w_0}=1$, and $\Delta_{t_0}=1$. Comparing with Fig.2, the function $P_{\Delta_{\theta_0}}(\theta-\theta_0)$ center around 0, which deviates more greatly form the expected value $\theta_0=\pi/4$. }
 \end{figure}
 \begin{figure}[h]
\includegraphics[scale=1]{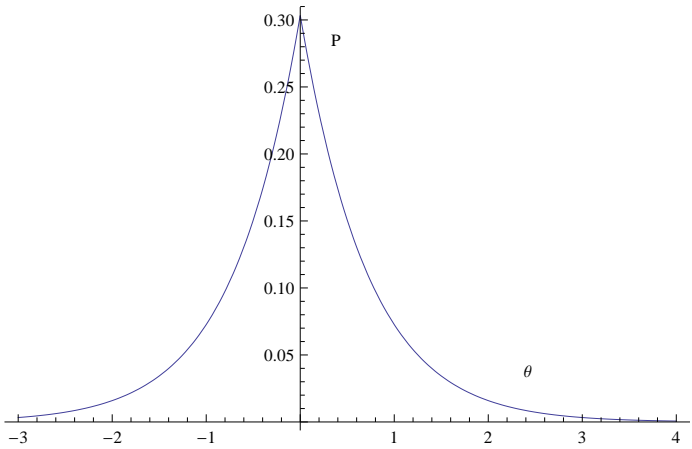}
\caption{\label{fig.2}The graph shows the continuous situation for the distribution. Here the parameter is given as: $w_0=1$, $t_0=0$, $\Delta_{w_0}=1$, and $\Delta_{t_0}=1$. }
 \end{figure}
 \section{decoherence environment}
 Any system inevitably suffers from the surrounding environment. We consider that during the rotation,  decoherence environment has an effect on the system.

 Let us consider that a classical environment (for example, random telegraph noise) contacts with the spin system. The interaction Hamiltonian can be written as $H_{int}=\beta(t)\sigma_z$\cite{lab17,lab18} ($\hbar\equiv1$), where $\beta(t)=\pm\gamma$ with equal probability.

 We first discuss about the fuzziness of reference from coarsened rotation frequency.
Apply the interaction Hamiltonian $H_{int}$ and the rotation Hamiltonian $w\sigma_x$ to the projection operator $O=\sigma_z$,
\begin{equation}
\begin{split}
&O_{\Delta_{\theta_0}}(w_0t)=\int dw P_{\Delta_{w_0}}(w-w_0)\\
&[\exp[i(w\sigma_x+\beta(t)\sigma_z)t]\sigma_z[\exp[-i(w\sigma_x+\beta(t)\sigma_z)t].
\end{split}
\end{equation}
\begin{figure}[h]
\includegraphics[scale=1]{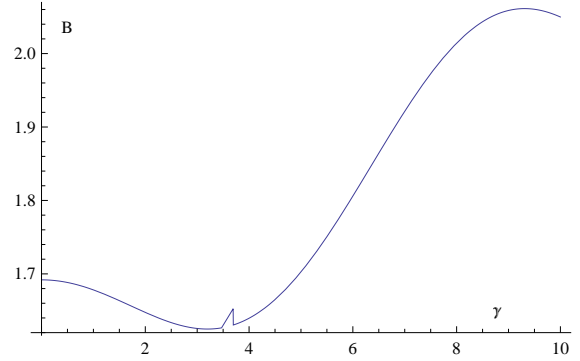}
\caption{\label{fig.5}The graph shows correlation between the Bell function $B$ and the coupling strength $\gamma$ where we only consider system $a$ suffering from the decoherence environment. Here the parameter is given as: $\theta_a=-\pi/8$, $\theta_b=0$, $\theta'_a=\pi/8$, $\theta'_b=\pi/2$, $\Delta_{w_0}=1$, and $|w_0|=1$. }
 \end{figure}
 \begin{figure}[h]
\includegraphics[scale=1]{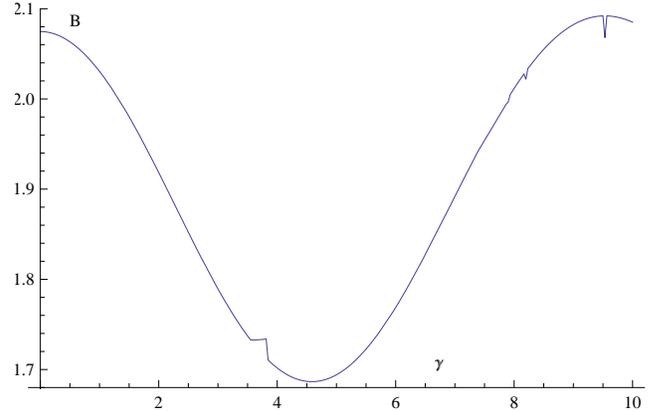}
\caption{\label{fig.6}Similar with the Fig.5, only reduce the standard deviations: $\Delta_{w_0}=0.5$, and $|w_0|=0.5$. We find that the decoherence environment can help to increase the value of $B$, when $B>2$ without decoherence environment ($\gamma=0$).}
 \end{figure}

\begin{figure}[h]
\includegraphics[scale=1]{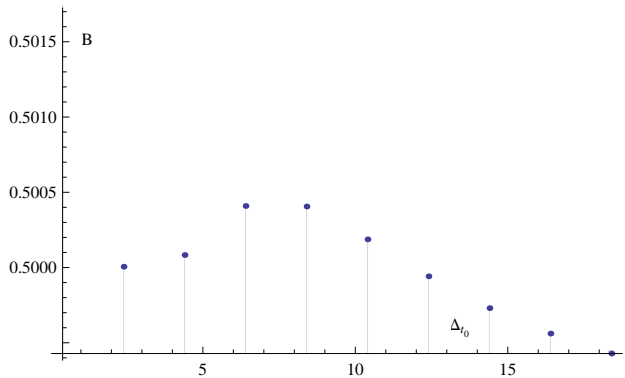}
\caption{\label{fig.7}The relationship between the Bell function $B$ and $\Delta_{t_0}$ is shown, where the parameter is given as: $\theta_a=0$, $\theta_b=7\pi/8$, $\theta'_a=\pi/4$, $\theta'_b=\pi/8$, $\gamma=1$, and $|w_0|=1$. }
 \end{figure}
As shown in Fig.5 and Fig.6, the decoherence environment can help to detect the quantum effects. Intuitively, the cause is that the decoherence environment changes the direction of the measurement axis, not along the $X$ axis.

Then we consider the situation about the fuzziness of reference from coarsened timing. In the decoherence environment, we find an abnormal phenomenon: the Bell function $B$ can increase with the standard deviation $\Delta_{t_0}$.

 \section{conclusion}
 We explore the origination of coarsened reference, and attribute it to the fuzziness of rotation frequency and time. As a result, the degree of fuzziness in coarsening the reference increases with the rotation angle for the coarsened rotation frequency, otherwise it is unchanged. For the coarsened rotation frequency, the degree of fuzziness of reference decreases with the central frequency and steps of rotation. On the contrary, it increases for the coarsened rotation time. For both coarsened rotation frequency and time, the rotation angle distribution will not be Gaussian and center around the expected value, which could influence the value of Bell function. Finally we study a simple classical decoherence environment during the rotation. Counterintuitively, the decoherence can help to increase the value of Bell function for detecting quantum effect. It is interesting to research the complex environment and the Bell inequalities for many particles in the coarsened reference \cite{lab19,lab20}. We believe that this article will deepen the understanding of coarsened reference in the quantum-to-classical transition and  how to control
the measurement reference.
\section{Acknowledgement}
This work was supported by the National Natural Science Foundation of China under Grant No. 11375168.

 \end{document}